\documentclass{aa}

\def\lte{{\sc lte}}

\def\doepsf{1}
\def\lii{\ion{Li}{i}}

\def\nlte{{\sc nlte}}
\def\aap{A\&A}
\def\apj{ApJ}

\begin{document}
\ifnum\doepsf=1\input{epsf}\fi

\thesaurus{02.12.1; 06.01.1; 06.07.2; 06.16.2}
\title{The 671\,nm \lii\ line in solar granulation}

\author{Dan Kiselman}
\institute{The Royal Swedish Academy of Sciences, Stockholm Observatory,
       SE-133~36 {\sc Saltsj\"obaden}, Sweden}
\date{Received ; accepted}
\maketitle
\begin{abstract}
The astro\-physically very interesting
\lii~671\,nm line has been observed with high spatial resolution 
in solar granulation with the intention to diagnose departures from
local thermodynamic equilibrium (LTE) in the line formation.
The spectral feature is very weak, so this is also a test on the
limits of such observations.
The observations of the \lii\ line and other weak lines nearby 
are compared with the results
of synthetic line calculations in three-dimensional
granulation simulations. The dependence of line-centre velocities on
photospheric continuum brightness is well described by the simulations.
The observed equivalent width of the \lii\ line show an approximately
flat dependence on continuum brightness, contrary to the theoretical
LTE results. Detailed modelling of the line radiative transfer, with an
approximate inclusion of three-dimensional effects, gives a better
agreement with observations. 
The match is not perfect and various interesting
reasons for this are considered. However, the
possibility of systematic errors caused by the sensitivity of the
\lii\ equivalent width to continuum placement
calls for cautiousness in the conclusions.
\end{abstract}

\section{Introduction}
Stellar lithium abundances are the subject of many current
debates in astrophysics (e.g., Thorburn 1996, Crane 1995, 
Spite \& Pallavicini 1995).
This paper continues the study of \lii\ line formation
reported by Kiselman (1997), hereafter referred to as Paper\,I.  The
aim is to improve the understanding of lithium
abundance determinations of solar-type stars by investigating the
formation of lithium lines outside the classical realm of
plane-parallel homogeneous photospheres and local thermodynamic
equilibrium (\lte).  It is also
intended to test the idea put forward in Paper\,I 
that spatially resolved solar spectroscopy
can be used to test \nlte\ results in less model-dependent ways than
when just line profiles in integrated light are used. At the same time
it tests the limits of such observational work since the lines studied
are very weak.

Paper\,I was motivated by the pro\-posal of 
Kurucz (1995) that lithium abundances derived for extremely 
metal-poor solar-type stars
using 1D photospheric models and assuming \lte\ may be underestimated 
by as much as 1\,dex due to 3D \nlte\ effects. 
The idea was that the transparency of ``cold'' regions in metal-poor
photospheres would cause all lithium there to be ionised by
ultraviolet radiation from ``hot'' regions. This would lead to a very
weak line in the resulting mean spectrum, since the contribution to
the spectral line from the hot regions is already small.

No such large effect of granulation on abundance determinations has
yet been demonstrated by observations or detailed simulations.
The standard behaviour of lines over the {\em solar} granulation
pattern seems to 
be that lines -- irrespective of the ionisation stage --
get stronger in bright granules and weaker in darker
intergranular lanes (Steffen 1989, Kiselman 1994). Thus one can expect
that the influence on abundance ratios, which are derived from line
ratios, will not be excessively large (Holweger et al. 1990).

The investigation of \lii\ line formation in a 3D model of solar
granulation in Paper\,I demonstrated that the Kurucz mechanism 
at least does not work in solar granulation. 
This is also the conclusion of
Uitenbroek (1998) who used a different \nlte\ treatment but the same
kind of granulation model. 

As regards metal-poor stars, the final verdict
must of course wait until we know more about the inhomogeneity properties
of their photospheres. 
This paper will only concern the solar \lii\ lines 
and is organised as follows. First the simulations are
discussed in somewhat more detail than in Paper\,I. Then the solar observations
are described, and finally the observations are compared with the
simulation results.

\begin{figure}
\ifnum\doepsf=1\hspace*{.2cm}\epsfxsize=8cm\epsffile{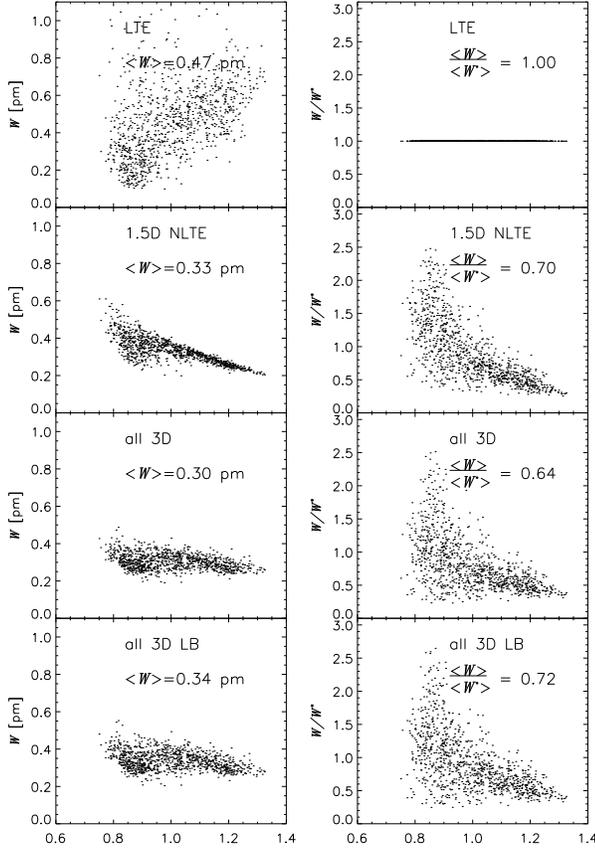}
 \else\picplace{1cm}\fi
\caption[]{Results from line transfer calculations in a granulation
snapshot using different treatments and approximations:
\lii\ equivalent widths as function
of continuum intensity (left) and the ratio of these to the corresponding
\lte\ values (right). Labels are explained in text.
The mean values given are intensity-weighted
means \label{fig_icwsol1}}
\end{figure}

\section{Simulations}
\subsection{Principles}
The availability of  computing methods, atomic
data, and powerful computers
may just begin to allow realistic and consistent solutions
of 3D \nlte\ problems in media such as photospheric granulation --
meaning large, and essentially 
complete, model atoms and full coupling between all angles. But 
such problems are still normally approached with some compromising
with regard to atomic models, radiative transfer, or geometry.
The aim of Paper~I was mainly to study the effects of \lte\
departures and 3D effects on the \lii\ lines
by comparing different spectral-line treatments in a single
solar-granulation-model snapshot, with standard 1D models only used as
a qualitative reference. The methods that were introduced, and are used also
here, involve the calculation of line
profiles and equivalent widths with a hybrid technique
using a 1D \nlte\ code and a 3D background radiation field
computed in \lte. This approach was chosen because it allowed a rather
simple application of existing codes and because 
the very low solar lithium abundance makes it
likely that the line 
radiative transfer problem can be treated as a continuum problem.

\subsection{3D photospheric models}
The 3D photospheric models are two snapshots from the
granulation simulations of Nordlund \& Stein (e.g., Stein \& Nordlund
1989, Nordlund \& Stein 1991) that were also
used by Kiselman \& Nordlund (1995) -- see that paper for
illustrations. The snapshots are chosen so that
they represent opposite phases in the (internally excited)
oscillation that is present in the
simulations. The original simulation data consisted of a $125 \times
125 \times 82 $ grid of thermodynamic quantities. These were cut and
resampled to a $64\times 64 \times 55$ grid, corresponding to
$6 \times 6 \times 1$\,Mm on the solar surface. The plots of
Fig.~\ref{fig_icwsol1} only show results for every second $x$ and $y$ grid
point, thus $1/4$ of the vertical columns. The snapshots are periodic
in the horizontal directions and this is used in the radiative
transfer calculations.

\subsection{\lte\ treatment of lines}
Several weak lines observed in the region around the \lii\ 671\,nm line are
also included in the analysis for reference. 
The lines which were found
in the VALD data base (Piskunov et al. 1995) or in the Kurucz (1995) line list were modelled
in \lte\ with the routines of Kiselman \& Nordlund (1995). The adopted
line parameters are given in Table~\ref{tab_lines}.

\subsection{Non-\lte\ \lii\ line transfer calculations}
The \lii\ line of interest is the resonance doublet at
671\,nm which is very weak in the solar spectrum due to the low
lithium abundance -- the standard value $A_{\rm Li} =
\lg {N({\rm Li})\over N({\rm H})} + 12 = 1.16$ 
(Grevesse et al. 1996; M{\"u}ller et al. 1975) is used here. 
The doublet is blended with several other weaker lines all of which are
not identified --  see Brault \& M{\"u}ller (1975) and Kurucz (1995). 
The stronger (shorter wavelength) doublet
component is less blended. The lines are also subject to hyperfine
splitting and isotopic splitting that cause complications when modelling
the line.

The \lii\ line profiles and equivalent widths 
were computed with version 2.2 of the plane-parallel \nlte\
code {\sc multi} (Carlsson 1986) which uses the operator perturbation
techniques of Scharmer \& Carlsson (1985).
All calculations of lines in emergent light 
were done for disk centre, i.e. for $\mu = 1.0$.

The 30-level lithium atomic model of Carlsson et al. (1994) was used
with one change:
the 671\,nm resonance line was treated as a single line and not as a
blended doublet. This was done because the current version of
{\sc multi} will not
treat the radiative transfer correctly when the (microscopic) line
profile is asymmetric in the presence of macroscopic velocity fields.
This treatment of the doublet as a single line will not cause errors
of importance for the current study since the line is very weak in
almost all cases discussed here, putting it
safely on the linear part of its curve of growth. This is confirmed
by the fact that the lithium abundance can be increased a factor of ten without
changing the qualitative behaviour of the line. The change in
line-formation height is apparently not significant for the \nlte\ behaviour.
Thus the equivalent
width of each line component will be proportional to the doublet
equivalent width. The absolute solar Li abundance will not be a central issue
here and line strengths will frequently be rescaled.

The option of providing {\sc multi} with a background radiation field for 
calculation of the photoionisation rates was used in all calculations
that are
presented here and in some cases also 
for bound-bound transitions. The background
radiation field was computed using routines written in the {\sc idl}
data processing language that are similar to
those used by Kiselman \& Nordlund (1995). The equation
of transfer was solved along a set of rays -- five inclination 
angles ($\cos \theta = \mu$) and eight azimuthal angles ($\phi$)
-- and the resulting intensities along each ray were used to
calculate a mean intensity $J_\nu$ in each $(x,y,z)$ point.
This was done with a pure Planckian source function without any 
allowance for scattering (i.e. in what is sometimes called ``strong \lte'').
Some interpolation
problems appeared close to $\tau_c = 1$ where the horizontal gradients
are strong. This caused spurious values of $J_\nu$ in a few points
but too deep down to have any impact on the resulting line profiles.

The hybrid treatment can make one worry about inconsistencies.
It is important that the opacities used for production of the
granulation model, the background radiation field, and the non-\lte\
calculations are compatible since a small difference may cause the
optical surface to fall at different depths with large effects on the
resulting outgoing intensities. Direct comparisons of the opacities
showed that the differences were indeed
insignificant within the context of the current study.

The presence of large amounts
of spectral lines, especially in the blue and ultraviolet,
is known to cause significant depression of the fluxes in these
spectral regions. To somewhat allow for this, the {\sc osmarcs}
plane-parallel \lte\ model-photosphere code (Edvardsson et al. 1993) 
was used to produce a solar model from which flux-weighted
mean opacities were computed for a wavelength range around the
wavelengths used in the computation of photoionising radiation fields.
The ratio between these mean opacities, including spectral-line
blocking, and the continuous opacities was then used to correct
the opacities in the computation via interpolation in temperature.
The correction factors range from 1.0 at high temperatures in the
deep layers to well above 10 in the low-temperature regions.

\subsection{Collisions with neutral hydrogen}
The importance of inelastic collisions with neutral hydrogen atoms for
photospheric line formation has been debated (e.g., Steenbock \& Holweger
1984, Caccin et al. 1993, Lambert 1993). Such processes were not
included here. It could, however, be interesting to see if spatially
resolved spectroscopy would offer a possibility to decide this issue
and pin down the relevant cross sections. Tests where collisional
rates were included according to the recipe of Steenbock \& Holweger
(1984) showed, however, that the changes introduced by the additional
collisions were much too small in this case to allow any observational test.  

\subsection{Results \label{sec_results}}
The left-hand plots of Fig.~\ref{fig_icwsol1} shows the equivalent width $W$
of the \lii\ 671\,nm line as a function of continuum intensity $I_c$
for one of the snapshots. Each point in the plots corresponds to one ($x,y$)
column in the simulation snapshot. The plots in the right column show the
departure from the \lte\ result in the form of $W/W^*$ ratios, once
again as a function of continuum intensities. From top to bottom, the
plots show the results of line calculations with increasing
sophistication. This series illustrates the importance of
the departures from LTE caused by lateral photon exchange and the
impact of ultraviolet line blanketing.

The uppermost pair of plots ({\bf LTE}) 
shows the \lte\ result that was discussed
in some detail in Paper~I. Here the local temperature sets the line
source function and the line opacity according to the Saha and Boltzmann
equilibria. The result is that the line generally is strongest in the
continuum-bright regions but the $I_c - W$ relation shows a
significant scatter.

Leaving the \lte\ approximation gives the result in the second pair of
plots ({\bf 1.5D NLTE}). Here the line strengths have been calculated
in \nlte, so that each column of the snapshot is treated like a
plane-parallel model with horizontal photon exchange 
neglected. The result is a narrow dependence of equivalent width on
continuum intensity since now the atomic level populations are
 governed by a radiation
temperature that is set in the lower regions where the continuum is
formed. The experiments in Paper~I showed that it is the bound-bound
radiative transitions of \lii\ that matters, ultraviolet
overionisation has little importance for the departures from \lte.

The third row ({\bf all 3D}) shows the results when all lines have
been calculated
as fixed rates given by the 3D radiation field. Now the value of
the mean intensity $J_\nu$ in each point is influenced by horizontally
neighbouring regions and this gives a larger spread in the plot since
the continuum intensity on the plot axis is the intensity coming from
straight below.

The final modelling  ({\bf all 3D LB})
-- which is to be compared with observations later -- 
is represented by the lowest pair of plots in
Fig.~\ref{fig_icwsol1}. Here all transitions are treated as fixed with
the 3D radiation field and that field has been corrected for line
blanketing as described above. The effect of the schematic
line-blanketing correction is to increase the line strength somewhat
due to the damping effect this has on the line-pumping-induced
overionisation. The difference from the \lte\ plot is
still significant and it seems possible to discern between the two
cases observationally as will be tried in the next section.

Uitenbroek (1998) studied \lii\ line formation with a consistent 2D
\nlte\ treatment, a similar granulation
snapshot, but a smaller
model atom than the one of this paper. The results are qualitatively
similar to those presented here, but the discussion of them
differs somewhat in that ultraviolet overionisation is considered to be
an important mechanism contrary to what was argued in Paper~I.

\begin{figure}
\ifnum\doepsf=1\hspace*{.2cm}\epsfxsize=9cm\epsffile{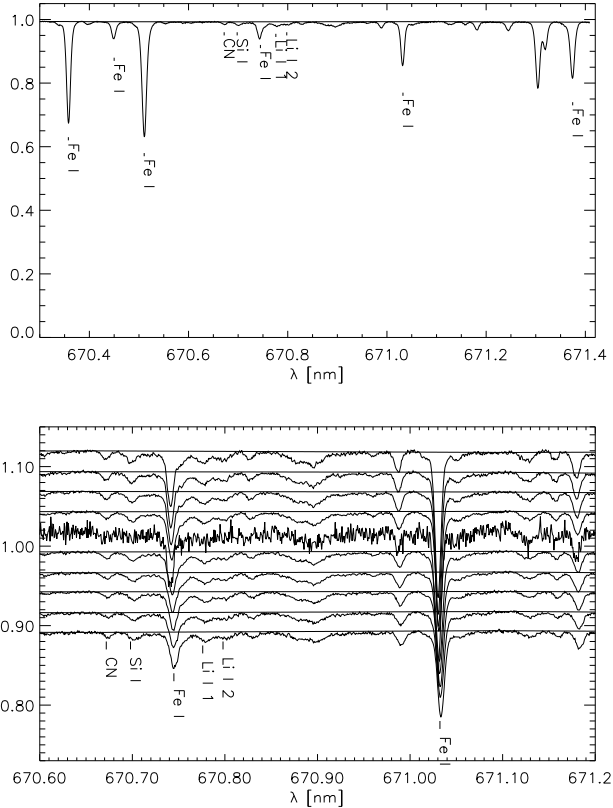}\else\picplace{1cm}\fi

\caption{The upper panel shows the mean spectrum from one exposure
with the lines that are included in the study marked. The lower panel
shows the binned spectra in the region around the \lii\ feature.
One single unbinned spectrum is included for reference. \lii\ 1 marks
the main doublet component which is the one that is measured. The
continua that are fitted to the spectra are shown \label{fig_spec}}
\end{figure}

\begin{table}
  \caption[]{Adopted quantities for the spectral lines and the
         compilations from which they were taken }
         \label{tab_lines}
      \[
         \begin{array}{llllll}
            \hline
            \noalign{\smallskip} 
                  &  \lambda~{[\mathrm{nm}]} & \chi_l~[\mathrm{eV}] & g_l & g_u &
         f \\ 
            \noalign{\smallskip}
            \hline
            \noalign{\smallskip}
            \ion{Fe}{i} & 670.36 & 2.76 & 1 & 1 & 6.90-4~^{\mathrm{a}} \\
            \ion{Fe}{i} & 670.45 & 4.22 & 1 & 1 & 2.19-3~^{\mathrm{a}} \\
            \ion{Fe}{i} & 670.51 & 4.61 & 1 & 1 & 3.19-2~^{\mathrm{a}} \\
            \ion{CN}{ } & 670.67 & 2.76 & 46 & 46 & 3.68-4~^{\mathrm{b}} \\
            \ion{Si}{i} & 670.70 & 5.95 & 1 & 1 & 3.31-3~^{\mathrm{b}} \\
            \ion{Fe}{i} & 670.74 & 6.45 & 5 & 5 & 5.89-2~^{\mathrm{c}} \\
            \ion{Li}{i} & 670.78 & 0.00 & 2 & 6 & 7.53-1~^{\mathrm{d}} \\
            \ion{Fe}{i} & 671.03 & 1.49 & 1 & 1 & 1.32-5~^{\mathrm{a}} \\
            \ion{Fe}{i} & 671.37 & 4.80 & 1 & 1 & 2.51-2~^{\mathrm{a}} \\
            \noalign{\smallskip}
            \hline
         \end{array}
      \]
\begin{list}{}{}
\item[$^{\mathrm{a}}$] VALD data base (Piskunov et al. 1995) 
\item[$^{\mathrm{b}}$] Kurucz (1995)
\item[$^{\mathrm{c}}$] Kurucz (1995), $f$ multiplied with 1000
\item[$^{\mathrm{d}}$] Carlsson et al. (1994)
\item[] The lines with $g_u =1$ and $g_l =1$ had only $gf$ values available.
\end{list}
   \end{table}

\section{Solar observations}
\subsection{Methods}
The weakness of the \lii\ feature in the solar spectrum makes
observations and analysis at high spatial resolution challenging.
Spectroscopic observations of the \lii\ resonance doublet at 671\,nm
in the Sun were made with the 0.48\,m Swedish Vacuum Solar
Telescope (SVST) on La Palma in October and November 1995.
The Littrow spectrograph described by Scharmer et al. (1985) was used
with an extra mask inserted in the
spectrograph housing to block reflections from the Littrow lens.
The spectrum was registered by a 10-bit Kodak Megaplus 1.6 CCD camera
operating synchronously with a similar camera that imaged the
slit-jaw. A spectrograph 
slit of 35\,$\mu$m width was used, corresponding to 0\farcs3.
This gives a nominal spectral resolution of 
$R = {\lambda \over \Delta \lambda} = 230000$.
All observations discussed here were made of quiet photosphere
at solar disk centre ($\mu = 1.0$) on November 2, 1997. The exposure time
was 150\,ms.

The first part of reductions were made with routines written for the
purpose in the {\sc ana} interactive data processing
language (Shine et al. 1988, Hurlburt et al. 1997).

Flat-fielding of the spectral frames
is a problem since there is no continuous light source
suitable to stand in for the Sun. The method introduced by Lites et
al. (1990) employs
the summing up of defocused exposures taken while the telescope scans
over the solar disk. A mean spectrum is constructed for each frame by
summing over the spatial direction.
The resulting spectral frames are then divided
with their mean spectra to get a flat field. The problem with this
procedure is that unwanted systematic vertical structures cannot be
removed and can even be created, e.g. by dust specks. To partially
take care of this problem, a reference spectrum was made using several
different grating settings so that each spectral region was recorded
using different parts of the detector. Comparing the reference
spectrum with the mean spectrum (from flat-field exposures) disclosed
low-spatial-frequency variations which could then be corrected for.
Apart from the flat-field and dark corrections, the spectra were corrected
for geometrical distortions and put on a common intensity scale with
the mean continuum intensity for an observational sequence as
reference.

After all these treatments and corrections, close inspection of
spectral frames, with the spectral lines and the continuum variations
removed by division with a continuum mask and the mean spectrum,
still showed a low-frequency fringe pattern with an
amplitude of about 0.5\,\%. This is the same order of magnitude
as the relative depth of the lithium line. The fringes was removed by
division with a sine-wave pattern whose parameters were found 
by a combination of least-square fitting and eye estimates. 
At the same time, the spectral frames were corrected to have the same
choice of continuum level as that of Brault \& M{\"u}ller (1975).

The resulting spectra differ from the output of a perfect instrument
due to image degradation by the atmosphere and the telescope, 
and scattered light in the
spectrograph after the slit. The impact of the former may be assessed
by measuring the continuum contrast in the spectra and looking at the
simultaneous slit-jaw image, the importance
of the latter can be deduced by comparing strong spectral lines with
a solar atlas of high spectral resolution and purity.
The strong Ca\,{\sc i} line at 672\,nm was observed to have a residual
intensity of 0.41 of the continuum level compared to 0.35 in the
digital issue of the solar atlas of Delbouille et
al. (1973). If this difference is interpreted as the result of the
addition of uniform stray light, the level of this stray light is
about 10\,\% of the ``true'' continuum level. This is apparently not unusual
for single-pass instruments  though it seems that
to diagnose the origins of such straylight 
can be difficult (Gulliver et al. 1996).

Seventeen of the best spectra were chosen for further analysis (from now on
using routines written in the {\sc idl} data processing language), 
they consist of $1523 \times 879$ arrays
corresponding to $1.1\,{\rm nm} \times 1\arcmin12\arcsec$. The pixelsizes in the
spectral and the spatial direction oversamples the
diffraction-limited spatial resolution ($\sim 0\farcs3$) and the
nominal spectral resolution ($\sim 3\,{\rm pm}$). The {\sc rms}
intensity contrast of
the spectra, as measured along the full length of the slit in a
continuum spectral region, is around 6\,\%. This is a quality measure
on the spectra that shows that the current observations do not rival
the best ever achieved with this instrument, probably those of Lites
et al. (1989), but are still good.

It is not possible to examine the \lii\ feature in each spectral row
since the S/N there is not high enough to allow reliable continuum
determinations for such weak lines. Thus, the spectra were binned
for each frame according to continuum intensity so that the variation
of line properties with photospheric brightness may be investigated.
For sufficiently strong and unblended lines, simple integration,
without any assumption on line shape,
is the best way to get equivalent widths.
For the weakest lines, a Gaussian profile was fitted to the central
part of the line feature.
The equivalent width of the \lii\ feature measured
in this way -- the profile is fitted to the strongest and least
blended doublet component -- 
will not be equal to the true equivalent width for the whole doublet.
 But since the line is on the linear part of
its curve of growth, the measured value will be closely proportional
to the ``real'' equivalent width which corresponds to what we get from
theory.
Figure~\ref{fig_spec} shows the binned spectra for one frame together
with one single spectral row.

In the further analysis of the data, the binned spectra with less than
20 single contributing spectra were discarded in order to decrease the
noise. These represent the extremes of the continuum intensity range.

\begin{figure*}
\ifnum\doepsf=1\hspace*{.2cm}\epsfxsize=17cm\epsffile{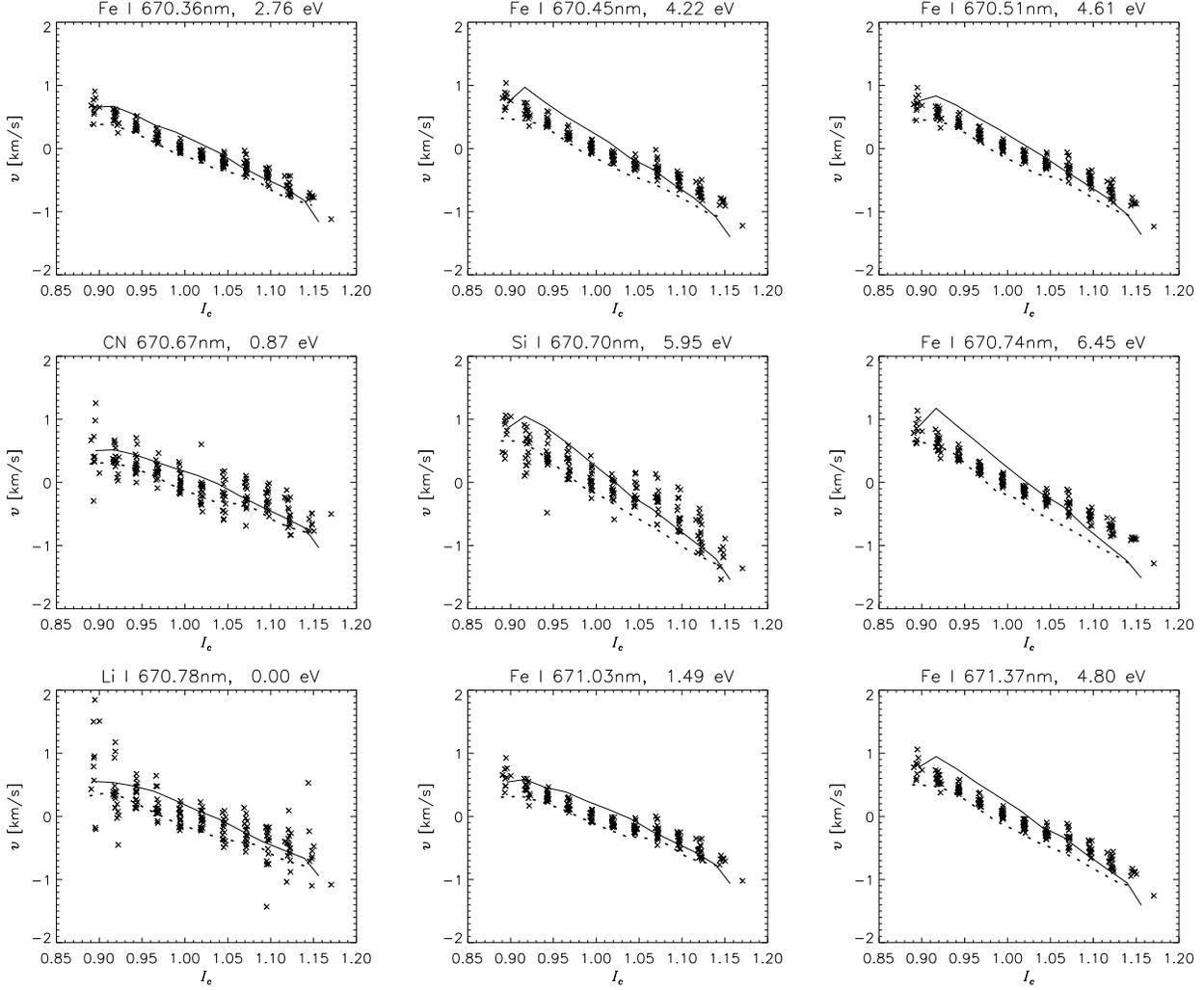}\else\picplace{1cm}\fi
\caption{Comparison of line-centre velocities $v$ and their dependence on
 continuum intensity $I_c$ as measured from binned
 observed spectra (crosses) with \lte\ simulations in two granulation
 snapshots (dashed and solid lines) for several weak
 spectral lines in the same spectral region as \lii  
 \label{fig_all_lincen}}
\end{figure*}

\begin{figure*}
\ifnum\doepsf=1\hspace*{.2cm}\epsfxsize=17cm\epsffile{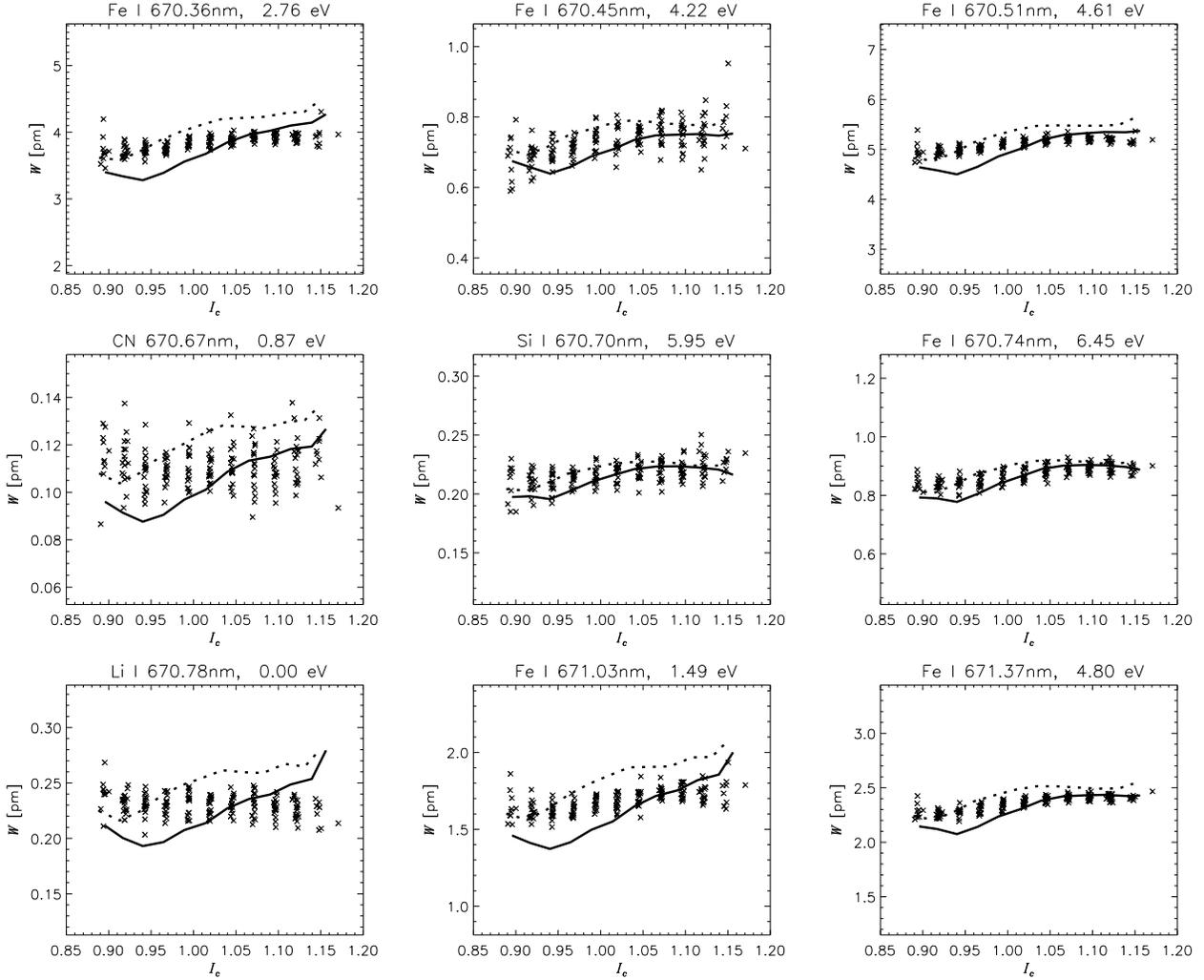}\else\picplace{1cm}\fi
\caption{Comparison of equivalent widths $W$ measured from 
 binned observations (crosses) with \lte\
 simulations in two granulations snapshots (dashed and solid lines) for
 several weak spectral lines in the same spectral region as \lii
 \label{fig_all_ww}}
\end{figure*}

\begin{figure}
\ifnum\doepsf=1\hspace*{.2cm}\epsfxsize=9cm\epsffile{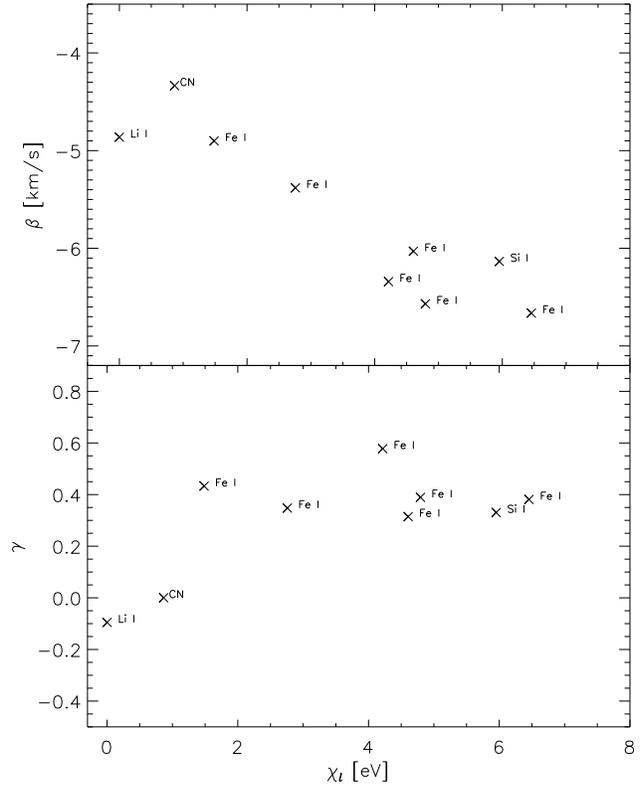}\else\picplace{1cm}\fi
\caption{ Plots showing the sensitivity on continuum intensity
   of observed line parameters to the excitation energies of the
   lines' lower levels. $\beta  = {dv\over dI_c} \cdot \langle
   I_c\rangle$, the slope of the
   line-centre velocities. $\gamma = {dW\over dI_c}
   \cdot {\langle I_c\rangle\over \langle W\rangle}$, the normalised
   slope of the equivalent widths.
 \label{fig_fincalc}}
\end{figure}

\begin{figure}
\ifnum\doepsf=1\hspace*{.2cm}\epsfxsize=9cm\epsffile{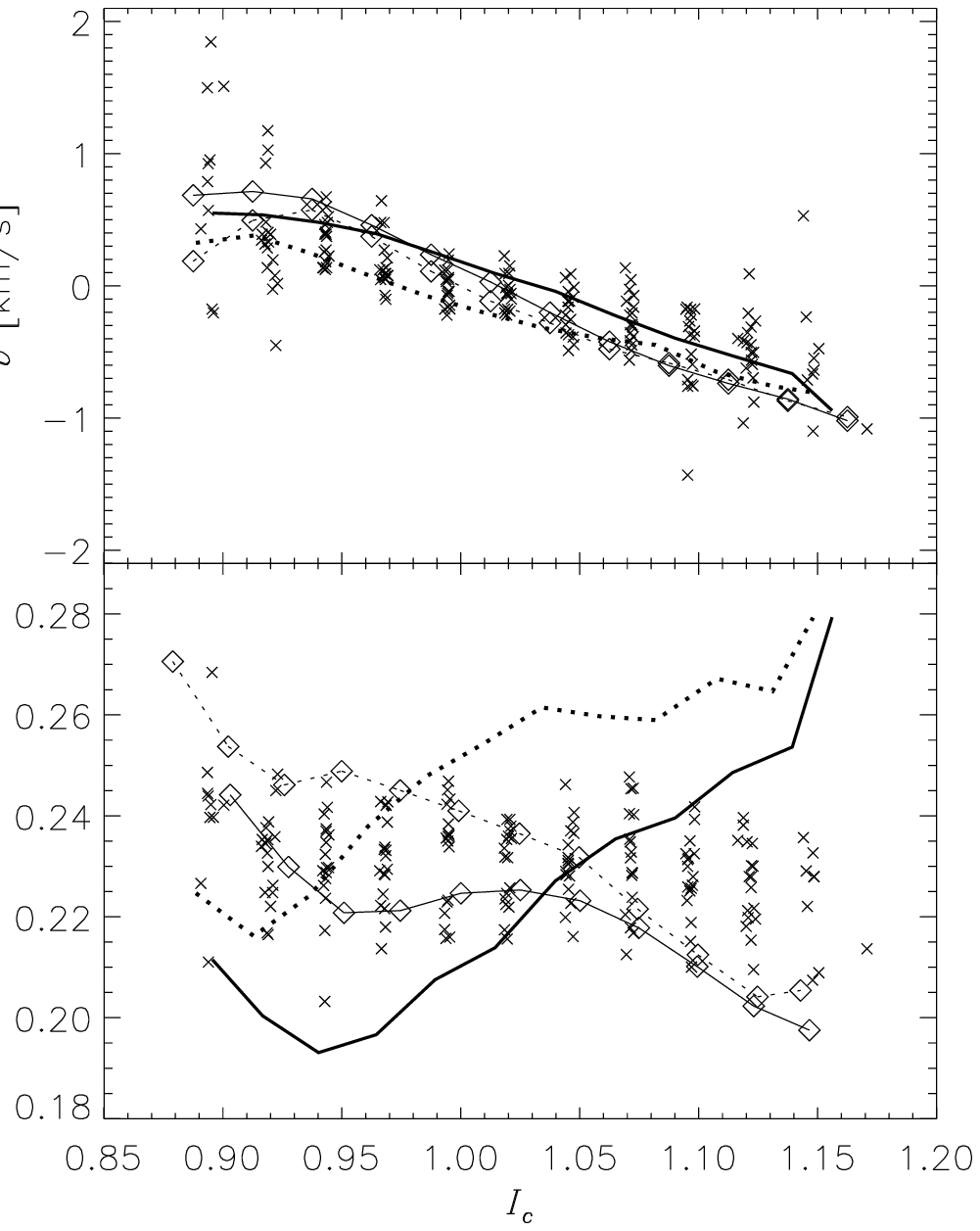}\else\picplace{1cm}\fi
\caption{Comparison of line-centre velocities (upper panel) and 
 equivalent widths (lower panel) of the \lii\ 671\,nm line 
 measured from binned observational
 spectra (crosses) with \lte\ and \nlte\ simulations of the
 \lii\,671\,nm line. Each cross comes from one of seventeen
 exposures. Thick dotted and full-drawn lines are \lte\
 simulations for the two granulations snapshots, 
 lines with diamonds are \nlte. The theoretical values
 have been adjusted so that their mean values coincide with that of
 the observations 
 \label{fig_finli}}
\end{figure}

\section{Comparison of observations with simulations}
\subsection{Smearing and binning of the simulations}
In order to allow a fair comparison between simulations and observations, the
simulated results must be treated in a way that mimics the effects of
the atmosphere, the telescope, and the spectrograph. For this study, a
simple model using the convolution with two Gaussians, one narrow and
one wider with lower amplitude, was chosen. The {\sc FWHM}:s and the
relative amplitudes of the Gaussians were chosen through experiments. The
intensities, equivalent widths, and line-centre positions from the
simulation were smoothed with various combinations of parameters until
a combination was found that gave reasonable {\sc RMS}-contrasts and the
simulated intensity image looked similar to slit-jaw images.
The {\sc FWHM} of the narrow Gaussian corresponds to 0\farcs5 and that of the
broader curve, with an amplitude of 0.25 of the narrow one, to 4\farcs3.

 The experiments showed that the basic feature used
here as a diagnostic, the $I_c - W$ diagram, did not change
dramatically in shape when
smoothed in different ways. Thus the detailed smoothing procedure used
is not crucial to the results (though not unimportant).

Finally, the simulation results were binned in the same way as the
observational spectra. Since the simulated data is noiseless, 
the lines that show simulation results in
the figures represent the mean quantities for each bin.

\subsection{Discussion}
\subsubsection{All lines}
Figure~\ref{fig_all_lincen} shows the observed and predicted
line-centre velocities as function of continuum intensity, measured
from binned spectra. The theoretical data come from the \lte\
simulations. A constant has been added to the velocities so 
that the intensity-weighted mean velocities for all data sets are zero.
The correspondence is generally good, giving some
confidence in the methods and models used here.

Figure~\ref{fig_all_ww} shows similar plots for equivalent widths. The
theoretical values have been multiplied by a constant so that their
intensity-weighted mean values coincide with that of the observations.
The observed equivalent widths (binned) are well described
by the simulations for some lines but not for others.
The Fe\,{\sc i} lines seem to be reasonably well described by the \lte\
modelling though with some deviations. The behaviour of the
Si\,{\sc i} line is well reproduced. The CN line shows what
appears to be a discrepant behaviour, but the weakness of the line
calls for cautiousness in the interpretation. The discrepancy between
observations and \lte\ theory for the \lii\ seems more significant.
The observed $I_c - W$ relation shows a weak slope in the opposite
direction to the predicted \lte\ curves.

The central idea of this paper is that discrepancies between the
theoretical \lte\ results and the observations of
Fig.~\ref{fig_all_ww} are due to \nlte\ effects. Such conclusions are
valid {\em if} we can trust the granulation models and the observational
data. The following discussion will concentrate on the \lii\ line
because of the (purported) 
reliability of the \nlte\ modelling for this light atom,
which can be represented with an essentially complete atomic model.
For heavier atoms like Fe\,{\sc i}, the problem of
assembling sufficiently complete and precise atomic data
makes \nlte\ spectral-line modelling difficult and always
questionable. The \lii\ line is thus
a good test case in theory, but its weakness introduces observational
uncertainties.

\subsubsection{Excitation-potential dependence}
The different slopes in the diagrams just discussed are
highlighted in Fig.~\ref{fig_fincalc} where the normalised slopes from the
observational data are
plotted against the lower-level excitation energies
of the respective lines. The normalised measures of the slopes are
defined as $\beta = {dv\over dI_c} \cdot {\langle I_c\rangle}$ for
the line-centre velocities and $\gamma = {dW\over dI_c} \cdot
{\langle I_c\rangle \over \langle W\rangle}$ for the equivalent
widths. The derivatives come from 
second-degree polynomial fits to the
observational points and their values at mean intensity are given.

The upper plot of Fig.~\ref{fig_fincalc} shows that the line-core
velocity slopes correlate well with excitation potential. This is
most easily understood as caused by the high-excitation lines being
formed at greater depths (and temperatures) where the vertical velocities are
greater than higher up.

The equivalent-width slopes plotted in the lower plot do not show a
clear correlation with excitation potential. The Fe\,{\sc i} lines
show slopes that do not increase for high-excitation lines.
This is similar to what was found for the Fe\,{\sc i}
lines in Figure~5 of Kiselman (1994), and could be a manifestation of
\nlte\ effects.

\subsubsection{The \lii\ line}
Figure~\ref{fig_finli} compares the observational 
line-centre-velocity and 
equivalent-width results for \lii\ with the theoretical predictions. 
The \lte\ results are the same as those in Figs.~\ref{fig_all_lincen}
and \ref{fig_all_ww}, the \nlte\ results are those labelled ({\bf all 3D
LB}) in Sect. \ref{sec_results} and Fig.~\ref{fig_icwsol1}.

The difference seen before between the
theoretical \lte\ and \nlte\ equivalent-width results
is now reduced because of the 
binning and smoothing operations. The spread among the observational points
(crosses) at a specific intensity value is probably mostly
due to random errors, though there are also inherent differences from
frame to frame. Note how this spread increases towards the limits of
the plots due to fewer points from very bright or very dark regions
contributing to each binned spectrum. (Remember that results from bins
with fewer than 20 contributors are not shown.)
In essence the observed equivalent widths
 fall in between the \lte\ and the \nlte\
curves, though closer to the latter. The \nlte\ curves fall inside the
range of the observational points for most of the intensity range, 
but this is not the clear confirmation of the \nlte\
results one could have hoped for.
The possible explanations for the discrepancy between observations and
simulations can be summarised as follows.

{\em The \nlte\ treatment.} The discrepancy could be
explained by underestimated collisional cross sections. This
explanation would be the natural choice according to the central
hypothesis of this paper. If we believe in the granulation models and
the observational results, the most important (or uncertain) 
cross sections could in 
principle be determined by tuning them until there is correspondence
between simulations and observations.
The \nlte\ results are sensitive to the input $J_\nu$ values, which are
computed in a simplified way with a coarse angular resolution.
The discrepancy could be explained
if $J_\nu$ in the visual region is underestimated in the dark regions
relative to the bright ones. There is, however, no evident reason for
this to be the case.

{\em The granulation snapshots.} While these kind of granulation
simulations have demonstrated their realism, one must
remember that many
approximations have been employed in their computation -- e.g. the
finite spatial resolution that precludes the small structures which
are certainly there in the real Sun.
Possibly the two snapshots used are unrepresentative, but it should be
noted that they are chosen to be independent and of opposite
oscillatory phases. Also the apparent success in the \lte\ modelling
(of both line strengths and line-centre
velocities)
for most of the other weak lines  observed here is evidence for the
snapshots being essentially realistic. Another piece of evidence for
this is the
general result that the (\nlte) \lii\ line strength depends more on the
temperature structure at greater depths than on the kinetic temperature in the
higher layers where the uncertainty of the granulation simulations can
be expected to be greatest.

 {\em The smearing procedure.} The smoothing of the simulations
  is made in a rather schematic way, but as discussed earlier, 
 this is not likely to cause significant changes in the $I_c -W$ diagrams.

 {\em Blends.} The \lii\ feature is contaminated by CN lines, as
      demonstrated by Brault \& M\"uller (1975). According to their
      analysis, the contribution is relatively small. It should,
      however, make the measured $I_c -W$ dependence flatter according
      to the behaviour of the CN line in Fig.~\ref{fig_all_ww}. If
      there is contamination of some other line, or if the CN
      contribution was much underestimated by Brault \& M\"uller, this
      could explain the observed behaviour. This would mean that the
      solar lithium abundance is significantly lower than previously
      thought. That would seem to be at odds with the behaviour of the
      \lii\ line in sunspots (e.g. Barrado y Navascu\'es et al. 1996).

 {\em Systematic errors in the measured equivalent widths.}  These are
      difficult to estimate. A direct integration gives insignificant
      differences compared to the Gaussian profile fits, so the latter
      procedure is probably
      not a major source of error. In the same way, making the
      experiment of subtracting 10\,\% of the mean light level from
      the spectral frames did not result in significant changes in the
      $I_c - W$ diagrams. Straylight is therefore not likely as an
      explanation for the discrepancy.
      The continuum placement is, however,
      crucial and expected to be the most important source of errors in the
      observational equivalent widths. An erroneously placed continuum
      will lead to an error in the $I_c - W$ slope that is not
      corrected when equivalent widths are rescaled with a constant
      factor. This was clear from the preliminary reduction work, though
      it should be noted that the algorithm for continuum fitting was
      chosen on what was considered to be 
      objective grounds with the continuum level placed close
      to that of Brault \& M{\"u}ller (1975).
      The size of these errors is very difficult to
      estimate, but there is a clear possibility that they {\em could} be
      significant.

Were it not for the possibility of observational errors due to the
continuum placement, the discussion would leave errors in the atomic data 
as the most probable explanation for the
discrepancy. Problems with the granulation simulations, with the
\nlte\ treatment or the possibility of blends
would also be interesting to investigate further.
The likely presence of systematic errors due to continuum-placement
problems makes it, however, somewhat uncertain that the discrepancy is real and
significant.

\section{Conclusions}
The observed $I_c - W$ relation for the \lii\
671\,nm line is rather flat and falls
in between the theoretical \lte\ and the \nlte\ relations, though
closer to the latter. It seems that the results
exclude \lte\ as a line-formation hypothesis, something that is of
interest in principle, though of course not very surprising. The \nlte\
modelling of this paper is at best only marginally confirmed, however.
It is the extreme weakness of the \lii\ line that
makes definite conclusions difficult to
make. This is both because the necessary binning makes it impossible
to study the spread in the $I_c - W$ plot and because the equivalent
widths are sensitive to errors in the continuum
placement. Spectra with better spatial resolution and/or higher S/N would
probably allow firmer conclusions if they could be acquired.

The idea that spatially resolved
spectroscopy may be helpful in revealing \nlte\ effects is still
viable. It should be tested on somewhat stronger lines of other
elements, even though their \nlte\ modelling is intrinsically more complicated
and uncertain than for the relatively simple \lii\ case.
A coming paper (Kiselman, in preparation) will discuss similar solar
observations and possible departures from \lte\ of a large set of Fe lines of
different strengths.

\begin{acknowledgements}
I thank all staff of the SVST for assistance during the
observations, G{\"o}ran Scharmer also for comments on the
manuscript, Martin Asplund and Rob Rutten are thanked 
for discussions, and {\AA}ke Nordlund for help with the granulation models.
\end{acknowledgements}

\end{document}